\documentclass[%
twocolumn,
 amsmath,amssymb,
 aps,
 showpacs
]{revtex4-1}

\usepackage{amsmath}
\usepackage{mathrsfs}
\usepackage{graphicx}
\usepackage{dcolumn}
\usepackage{bm}
\usepackage{ulem}

\begin{document}
\title{High harmonic generation from pre-ionized H$_2$ in ultrashort intense laser fields
}

\author{Behnaz Buzari$^a$}
\author{Mohsen Vafaee$^{b}$}
\email{ E-mails: m.vafaee@modares.ac.ir}
\author{Hassan Sabzyan$^{a}$ }
\email{ E-mails: sabzyan@sci.ui.ac.ir}
\affiliation{
$^{a}$ Department of Chemistry, University of Isfahan, Isfahan 81746-73441, I. R. Iran
\\$^{b}$ Department of Chemistry, Tarbiat Modares University, P. O. Box 14115-175, Tehran, I. R. Iran}
\begin{abstract}

Effects of the laser pulse wavelength and intensity on the HHG production from the \textit{ionic} and \textit{homolytic} pre-ionization transient states of the two-electron H$_{2}$ system exposed to ultrashort intense laser pulses are studied by solving time-dependent Schr\"{o}dinger equation.
It is found that for the populated homolytic species in each half cycle of the laser pulse having enough strength, a pair of strong and weak HHG radiations are produced simultaneously. 
For the populated ionic species divided into two regions, in each half cycle, one of the ionic regions has a strong radiation and the other ionic region has a weak HHG radiation.
The HHG spectra of the homolytic and ionic species are almost similar, except that in some limited parts of the HHG spectrum, one of them dominate the other. 
\end{abstract}

\maketitle

\section{Introduction}
Laser-induced nonperturbative phenomena such as nonsequential and high-order harmonic generation (HHG) processes in atoms and molecules have formed a growing area of research \cite{Becker_2012_RMP,Chang_2011, Joachain_2012, Winterfeldt_2008_RMP, Krausz_2009_RMP}. Electron correlation constitutes a basic resource for the understanding of the dynamics of many-particle systems \cite{Becker_2006}. Possibility of monitoring molecular dynamics based on the high harmonic generation produced by coherent electrons \cite{6}, imaging the geometrical structure of the H$_2^+$ molecular ion \cite{7} and calculation of the fragmentation probabilities and kinetic-energy spectra for Coulomb explosion of H$_2^+$ molecular ion \cite{8} are all consequences of electron density evolution of molecules in intense external fields. 
The dependence of HHG on the molecular orientation \cite{Lein-Corso_2003, Bian-Bandrauk_2012}, electron entanglement and its internuclear distance dependency \cite{22}, laser-induced electron recollision \cite{23} and comparative angular distributions of the ionized H$_{2}$  molecules by the linear and circularly polarized intense ultrashort (few-cycles) XUV laser pulses are investigated based on the numerical solution of the time-dependent Schr\"{o}dinger equation (TDSE) \cite{24}  and the frozen core Hartree-Fock method \cite{19}.

Responses of many electron atoms and molecules exposed to both weak and strong monochromatic fields have been studied by Floquet perturbation theory \cite{10}. 
Manipulation of molecules in strong laser fields such as multiphoton excitation and ionization, and dynamical response of individual valence electrons of molecules to strong fields have also been investigated by time-dependent density-functional theory \cite{11,12,13,14} and numerical integration of TDSE without Born-Oppenheimer approximation \cite{15,16}. 
Some ab initio (TDSE) calculations of the interaction with intense few-cycle near-infrared laser pulses beyond the one-dimensional approximation have been reported by Parker et al. \cite{Parker_2006} and Ruiz et al. \cite{Becker_2006} for the helium atom and hydrogen molecule respectively. 
In the work reported by Ruiz et al. on the hydrogen molecule, the electron motion is included in its full dimensionality, while the center-of-mass motion is restricted along the polarization axis, and the system exhibited a rich double ionization quantum dynamics in the direction transversal to the field polarization \cite{Becker_2006}.	
However, most computations are carried out with some approximations in the dimensions of the electron, and in the nuclear dynamics \cite{Camiolo_2009,SVS_2011,VSSBS_2012,BVS_2012,Band_C_L_2012}. 
In these studies of the interaction of the intense laser pulse with the hydrogen molecule, the electronic and nuclear dynamics have been considered in one dimension approximated by soft-core potential and classical mechanics, respectively \cite{Camiolo_2009,SVS_2011,VSSBS_2012,BVS_2012}. In a recent work, the nuclear dynamics in one dimension has been treated quantum mechanically \cite{Band_C_L_2012}. 

In our recent works, details of the dynamics of electrons and nuclei in the hydrogen molecule have been studied \cite{VSSBS_2012,BVS_2012}. 
We introduced in \cite{VSSBS_2012} a new simulation box setup for the precise description of the evolution of two electronic systems in intense laser pulses based on which different regions of the hydrogen molecule are well discernible. 
In \cite{VSSBS_2012}, details of the pre-ionization hydrogen molecule are studied and the underlying mechanisms responsible for the formation and evolution of the homolytic and ionic transient species are explored.
We investigated in \cite{BVS_2012} contribution of the \textit{homolytic} $(e_1 H_\alpha^+-H_\beta^+ e_2 $ $\&$ $ e_2 H_\alpha^+-H_\beta^+ e_1 )$ and \textit{ionic} $(H_\alpha^+-H_\beta^- $ $\&$ $ H_\alpha^--H_\beta^+ )$ transient states in the two-electron evolution of the 1-D H$_2$-system.

In the present work, we have focused on the norm evolution and the time profile HHG produced from the homolytic and ionic species as whole subsystems of the H$_2$ system. 
The HHG time profiles and spectra produced by different transient (doorway) preionization species under different laser pulse wavelengths and intensities are calculated and analyzed comparatively. 
Effect of the motion of nuclei on this evolution and its consequent HHG are investigated semiclassically. 
In this report and in the following work \cite{SVANB_2013} effects of the intensity and frequency of the laser pulse on the H$_2$-ultrashort laser pulse interaction are investigated in order to develop realistic setup for prospective time-resolved experiments on two-electron systems exposed to ultrashort laser pulses.

This paper is organized as follows: In Sec. II, the TDSE used for the study of electron dynamics, and semi-classical approach adopted for the study of nuclear dynamics in the two-electron diatomic systems exposed to intense linearly polarized laser pulses are presented. 
Section III presents and discusses time-dependent behavior of electrons, and the effect of the motion of nuclei. Also, different characteristics of the time and frequency dependencies of the harmonic spectra produced by different species of the pre-ionization hydrogen molecule are presented and analyzed. Finally, conclusions drawn from this work are highlighted in Sec. IV. Throughout this paper, atomic units (a.u.), i.e. $e=1$, $\hbar=1$ and $m_e$=1, are used unless stated otherwise. 

\section{Computational Methods}
\begin{figure}
\begin{center}
\begin{tabular}{c}
\resizebox{50mm}{!}{\includegraphics{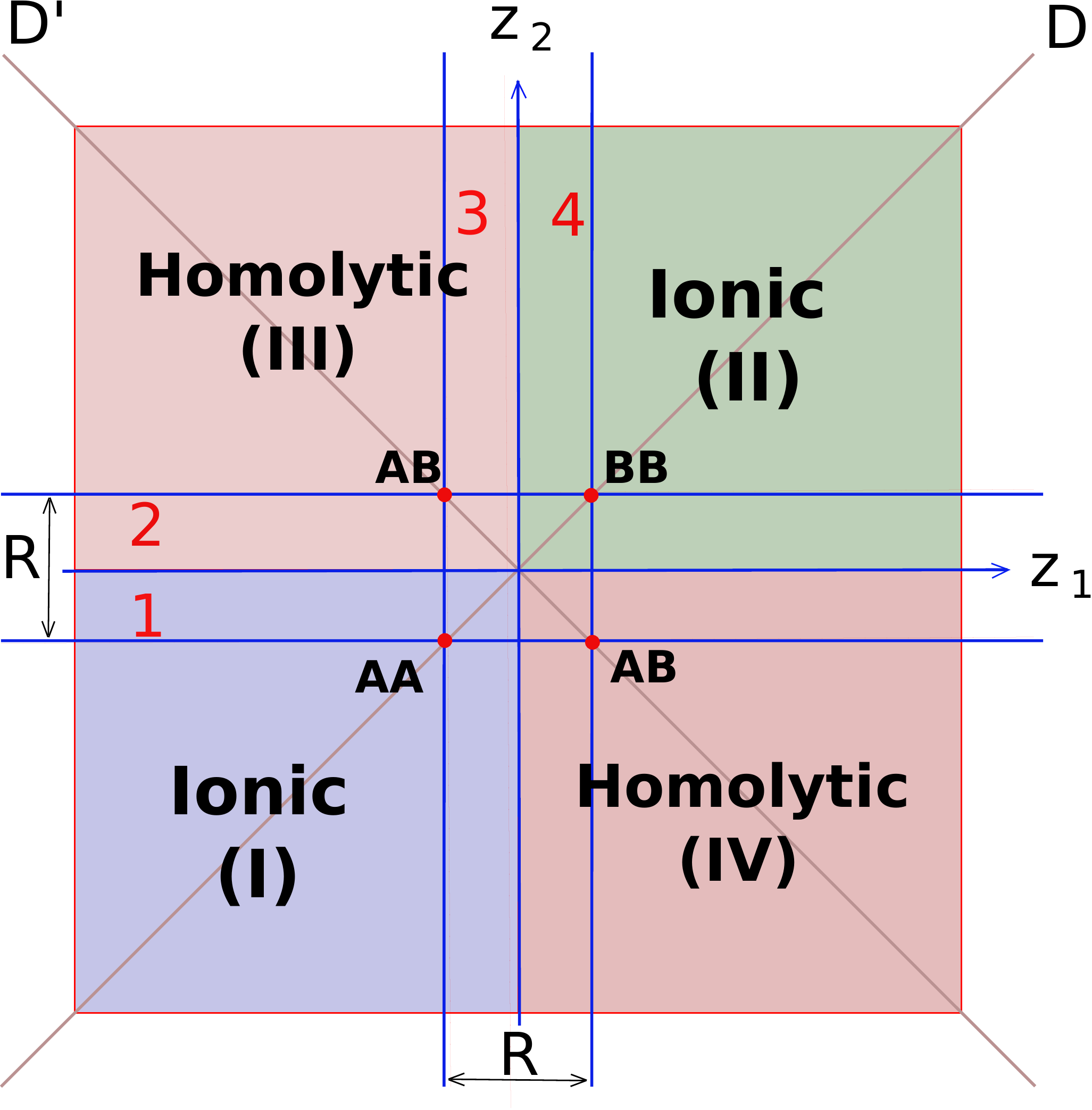}}
\end{tabular}
\caption{
\label{PIR}
(color on line) The pre-ionized regions of the two-electron molecular H$_2$; \textit{ionic I} $(H_\alpha^+-H_\beta^-$), \textit{ionic II} ($ H_\alpha^--H_\beta^+)$, \textit{homolytic III} $(e_1 H_\alpha^+-H_\beta^+ e_2$), and \textit{homolytic IV} ($ e_2 H_\alpha^+-H_\beta^+ e_1)$. See also Fig. 1 of Ref. [25] and Fig. 3 of Ref. [24]. }
\end{center}
\end{figure}

To investigate the electronic dynamics of the two-electron molecular H$_2$ in a strong laser field, we solve the TDSE for a 1D model with a soft-core Coulomb interaction between the charged particles. For this solution, we adopt a partitioning scheme for the simulation box which has been introduced and applied in our previous works  \cite{BVS_2012,VSSBS_2012}. This simulation box is shown in Fig.~\ref{PIR} (Fig. 3 of Ref. \cite{VSSBS_2012}).
In order to simplify the numerical solution of the electron-nuclear TDSE within the adiabatic approximation \cite{Camiolo_2009}
a semiclassical approach has been adopted to calculate the instantaneous forces applied on each nucleus by the other, the evolving electronic wavepacket and the laser field \cite{Camiolo_2009,SVS_2011}.
The instantaneous positions of the nuclei needed for the evaluation of the forces are calculated based on the Verlet algorithm \cite{Verlet}.

Assuming a linearly polarized laser pulse, and considering the fact that dynamics of the electrons and the nuclei occur in the laser field direction, we adopt a one-dimensional model for both the electrons and nuclei. In what follows, $R_1$ and $R_2$ indicate the nuclei positions and $z_1$ and $z_2$ are the electrons coordinates. Furthermore, $M$ and $m$ indicate the nucleus and electron masses, respectively, and $e$ is the electron charge. The temporal evolution of electrons of this system is described by the time dependent Schr\"{o}dinger equation (TDSE), i.e. \cite{Camiolo_2009,SVS_2011}
\begin{eqnarray}\label{eq:1}
&&i\frac{\partial \psi(z_1,z_2,t; R_1(t), R_2(t))}{\partial t}
=
\nonumber \\
&&
H_e(z_1,z_2,t; R_1(t), R_2(t)) \psi(z_1,z_2,t; R_1(t), R_2(t))
\end{eqnarray}
where the electronic Hamiltonian for this system, H$_e(z_1,z_2,t; R_1(t), R_2(t))$, is given by
\begin{eqnarray}
&&H_e(z_1,z_2,t; R_1(t), R_2(t)) =
\nonumber \\
&&-\frac{1}{2m_{e}} \left[\frac{\partial ^2}{\partial z_1^2}+\frac{\partial ^2}{\partial z_2^2}\right]
+V(z_1,z_2,t; R_1(t), R_2(t)),
\label{eq:2}
\end{eqnarray}
in which the potential V is
\begin{eqnarray}
&&V(z_1,z_2,t; R_1(t), R_2(t)) =
\nonumber \\
&& \sum_{i, \alpha=1}^2 \left(\frac{-Z_\alpha}{\sqrt{( z_i-R_\alpha)^2+a }}\right)
 +\frac{1}{\sqrt{( z_1-z_2)^2+b }}
 \nonumber \\
&&  +\frac{Z_1Z_2}{\sqrt{( R_1-R_2)^2+c }}+(z_1+z_2) \varepsilon(t),
\nonumber\\
\label{eq:3}
\end{eqnarray}
where $ Z_1=1$ and $Z_2=1$ are the charges of nuclei and
the screening parameters $a$, $b$ and $c$ are responsible for the softening of the electron-nuclei, electron-electron and nucleus-nucleus interactions, respectively. The values of these parameters are set to the same values as used by Camiolo et al. \cite{Camiolo_2009}.

The initial state is a singlet ground electronic state with an equilibrium internuclear distance R = 2.13 at rest. In this singlet state, the fermion electrons adopt an antisymmetric spin configuration, and thus the electronic spatial part of the wave-function $\psi(z1 , z2 , t; R1(t), R2 (t))$ is symmetric with respect to the permutation of the two electrons. The details of the numerical simulation is presented in Ref.~\cite{SVS_2011}).
Here in this work, linearly polarized laser pulses of $\lambda$= 532 and 390 nm wavelengths corresponding respectively to the fixed carrier frequencies of $\omega_{\circ}$=8.57$\times 10^{-2}$ and $\omega_{\circ}$=0.11683 a.u. with different intensities I=1$\times 10^{14}$, 5$\times  10^{14}$, 1$\times 10^{15}$ and  5$\times10^{15}\, Wcm^{-2}$ (i.e., $0.0028 $, 0.01.42, 0.028, 0.142 a.u., respectively) are used. 
For all cases, a pulse shape of $\varepsilon(t)=E_0 \, sin^{2} (t \pi/ \tau ) \, cos(\omega t)$ is used for the electric field of the laser, and the pulse length is set to $\tau$=8 cycles, along with a fixed time step of $\triangle t= 0.02$.
We set $d_{H_2}$, $d_{H_2^+}$ and $d_{H_2^2+}$ sizes (defined in Fig.~3 of Ref.~\cite{VSSBS_2012}) to 60, 500 and 580 (i.e. corresponding to the positions of the region boundaries at $\pm30$, $\pm250$ and $\pm290$, respectively) for $\lambda$=390 nm wavelength, and to 60, 1200 and 1280 (i.e. corresponding to the positions of the region boundaries at $\pm30$, $\pm600$ and $\pm640$, respectively) for $\lambda$=532 nm wavelength \cite{VSSBS_2012}.

Special attention is paid to the dynamics of individual \textit{homolytic} $(e_1 H_α^+-H_β^+ e_2 )$ and \textit{ionic} $(H_\alpha^--H_\beta^+ )$ quasi-states (doorways) formed transiently during the evolution of the two-electron wavepacket of H$_2$ based on the power spectrum of high harmonic generation (HHG) obtained from the mean value of the force acting on electron j (i.e. the electron acceleration) via \cite{Ban_Che_Lu_2013}
\begin{eqnarray}
\label{eq:3}
S_j(\omega)=\begin{vmatrix}\frac{1}{T_{tot}}\int_{0}^{T_{tot}}\langle \ddot{r}_j(t)\rangle exp(-\begin{scriptsize}\end{scriptsize}i \omega t)dt \end{vmatrix}^2,
\end{eqnarray}
and the time profiles of the high harmonics obtained via a Morlet wavelet transform \cite{Ban_Che_Lu_2013,Uzer_2003} of the time dependent dipole acceleration given by
\begin{eqnarray}
\label{eq:4}
&&W_j(\omega,t)=
\sqrt{\frac{\omega}{\pi^{\frac{1}{2}}\sigma}}\times
\nonumber\\
&& \int_{-\infty}^{\infty}\langle \ddot{r}_j(t')\rangle exp(-i \omega (t'-t))exp\left(\frac{-\omega^2 (t'-t)^2)}{2(\sigma)^2}\right) dt',
\nonumber \\
\end{eqnarray}
in which we set $\sigma$=2$\pi$.
\\

\section{Results and Discussion}
\begin{figure}[ht]
\begin{center}
\begin{tabular}{c}
\resizebox{90mm}{!}{\includegraphics{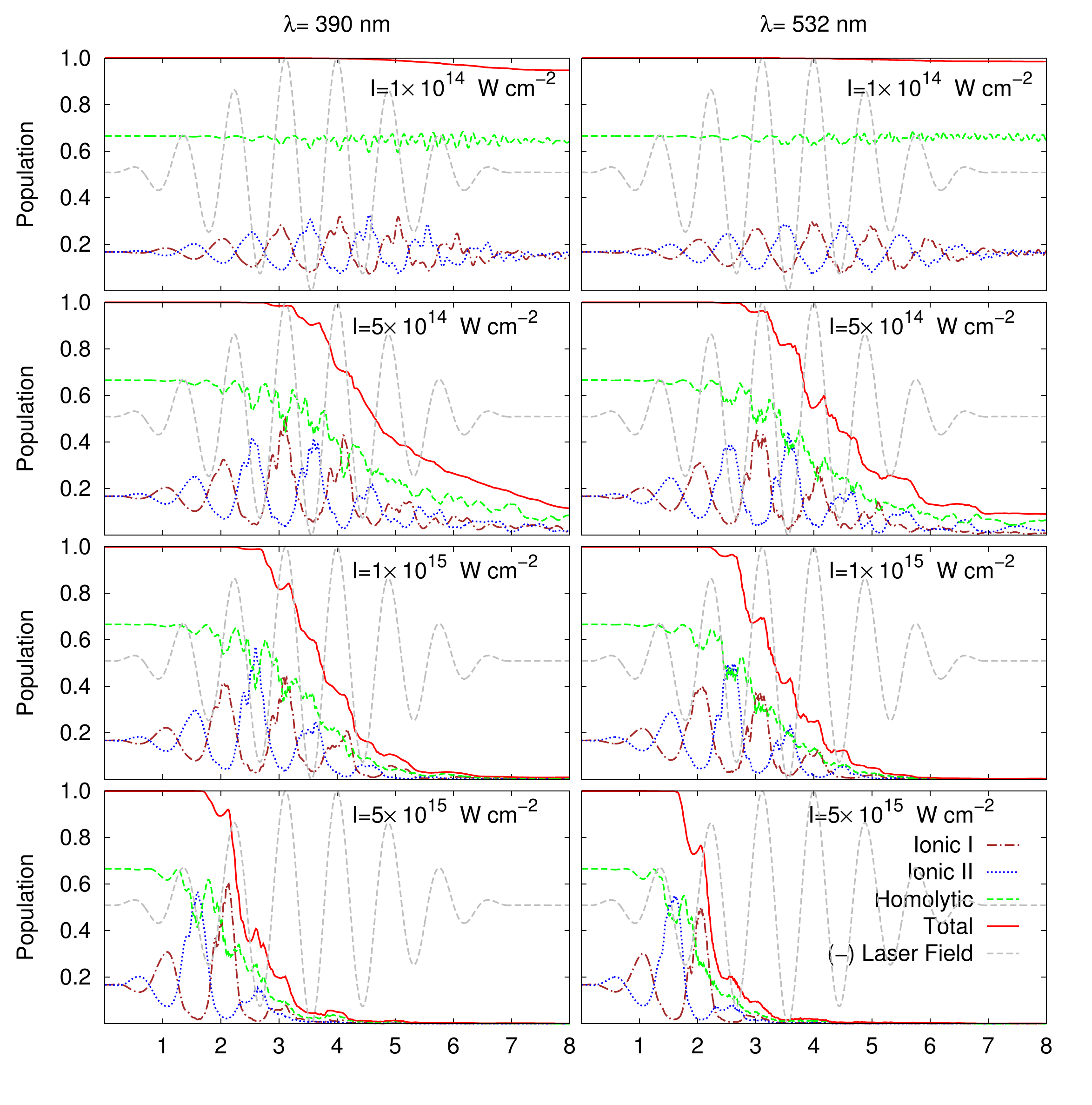}}
\end{tabular}
\caption{
\label{norm}
(color on line) Variation of the populations of the pre-ionization molecular H$_2$ (\textit{Total}) and transient 
\textit{ionic I}  $(H_\alpha^+-H_\beta^-)$, \textit{ionic II} $(H_\alpha^--H_\beta^+)$, and \textit{homolytic} $(e_1 H_\alpha^+-H_\beta^+ e_2 $ $\&$ $ e_2 H_\alpha^+-H_\beta^+ e_1 )$ species related to the H$_2$ region (Fig. 1) during the interaction of the H$_2$ system  with 8-cycle ultrafast intense laser pulses of $\lambda$= 390 nm (left) and 532 nm (right) wavelengths and at I=$1\times 10^{14}, 5\times10^{14}$, $1\times 10^{15}$ and $5\times 10^{15}\, Wcm^{-2}$ intensities. The same line style is used for all panels.
		}
\end{center}
\end{figure}

\begin{figure}[ht]
\begin{center}
\begin{tabular}{c}
\resizebox{50mm}{!}{\includegraphics{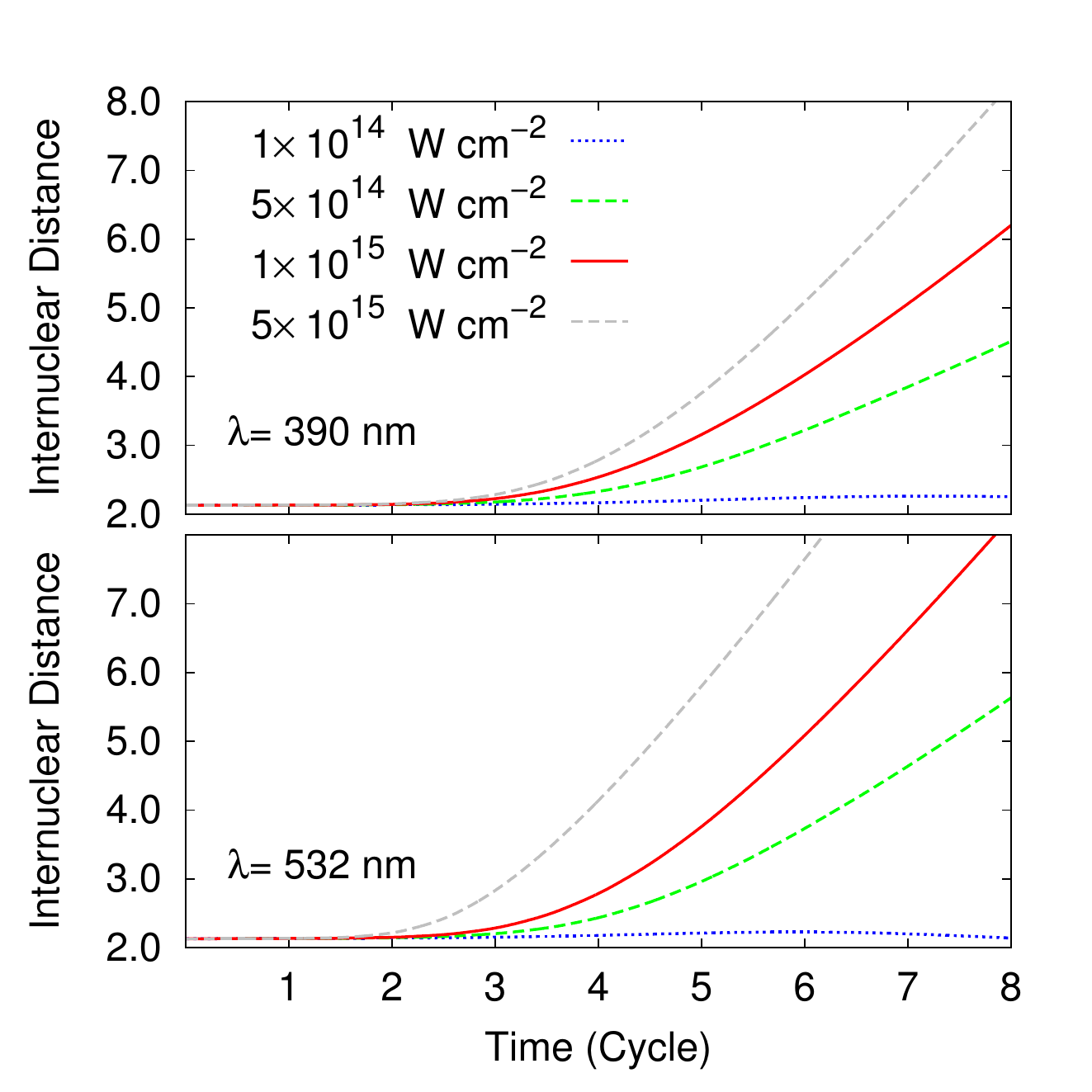}}
\end{tabular}
\caption{
\label{internuclear_distance}
(color on line) Internuclear distance of the H$_2$ system during the interaction  with 8-cycle ultrafast intense laser pulses of $\lambda$= 390 nm (left) and 532 nm (right) wavelengths and at I=$1\times 10^{14}, 5\times10^{14}$, $1\times 10^{15}$ and $5\times 10^{15}\, Wcm^{-2}$ intensities.
		}
\end{center}
\end{figure}

\begin{figure}[ht]
\begin{center}
\begin{tabular}{l}
\resizebox{90mm}{!}{\includegraphics{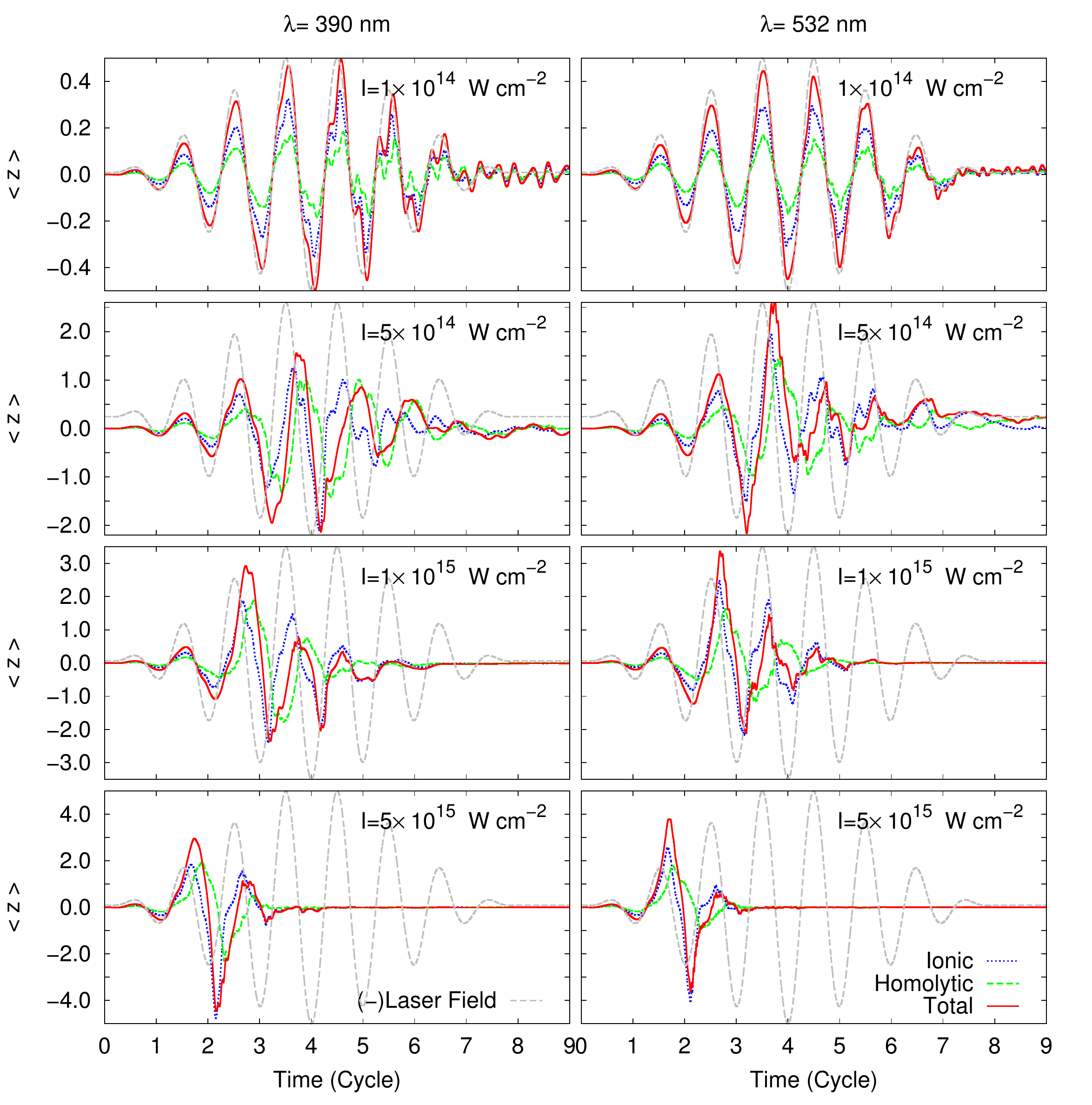}}
\end{tabular}
\caption{
\label{<z>}
(color on line) Variation of the expectation (average) values of electron position, $\langle z\rangle$, calculated for the pre-ionization molecular H$_2$ (\textit{Total}) and its transient \textit{ionic} $(H_\alpha^+-H_\beta^- $ $\&$ $ H_\alpha^--H_\beta^+)$ and \textit{homolytic}
 $(e_1 H_\alpha^+-H_\beta^+ e_2 $ $\&$ $ e_2 H_\alpha^+-H_\beta^+ e_1 )$ species during the interaction of the H$_2$ system with 8-cycle ultrafast intense laser pulses of $\lambda$ = 390 nm (left) and 532 nm (right) wavelengths and at I=$1\times10^{14}, 5\times10^{14}, 1\times10^{15}$ and $5\times10^{15}\, Wcm^{-2}$ intensities.
The same line style is used for all panels.		
		}
\end{center}
\end{figure}

Time-dependent populations of the \textit{ionic} and \textit{homolytic} species of the pre-ionized H$_2$, introduced in Fig. 1 of Ref \cite{BVS_2012} and summarized in Fig.~\ref{PIR} of  this report, during its interaction with laser pulses of different wavelengths and intensities are plotted in Fig.~\ref{norm}.
This Figure shows that the \textit{homolytic} pre-ionized species is the dominant transient state at earlier times of the course of interaction, and its contribution to the overall population decreases with time. 
A very similar variation and fall off is observed in the populations of these species in the calculation with the fixed nuclear distance (not reported here). This similarity is quite obvious if considering relatively small changes in the internuclear distance (Fig.~\ref{internuclear_distance}) in the corresponding laser pulse intensity and wavelength. 
Comparison of Fig.~\ref{norm} and Fig.~\ref{internuclear_distance} shows that the change in the internuclear distance is small while there is considerable population in the pre-ionization molecular H$_2$. 
It can be seen from Figs.~\ref{norm} and ~\ref {internuclear_distance} that the initial internuclear distance does not change significantly while there is considerable population in the pre-ionized H$_2$. 
Figure~\ref{norm} shows also 
that at the intensity of $I=1\times 10^{14}\, Wcm^{-2}$, the decrease in the population of the pre-ionized H$_2$ species is slightly larger for the $\lambda=390$ nm wavelength as compared to that for the  
$\lambda=532$ nm. While, at all other intensities, fall off in the population of the pre-ionized  species of H$_2$ is slightly steeper for the $\lambda=532$ nm than that for the  $\lambda=390$ nm wavelength. 
Variations of the average position $<\!z\!>$ of the electrons (1 or 2) of the \textit{homolytic} $(e_1 H_\alpha^+-H_\beta^+ e_2$ $\&$ $e_2 H_\alpha^+-H_\beta^+ e_1 )$ and \textit{ionic} $(H_\alpha^+-H_\beta^-$ $ \&$ $ H_\alpha^--H_\beta^+ )$ species in different laser pulse intensities and wavelengths are calculated and plotted in Fig.~\ref{<z>} as functions of time. This 
Figure shows that the ramp in the average position for both species becomes steeper and its oscillations decay off faster with increasing intensity.  
At the lowest laser pulse intensity, $I=1\times 10^{14}\, Wcm^{-2}$, oscillations of the electron position for both species follows exactly the oscillations of the laser pulse. 
The amplitude of these oscillations decreases with intensity which is due to the increase in the outgoing of the population reaching the simulation box borders of the pre-ionized H$_2$ region (Fig.~\ref{PIR}) as intensity increases. 
But, the time period of these oscillations increases with intensity which is due to further distancing of the electron population from the nuclei and more spreading over the simulation box as intensity increases.
It can also be seen that the homolytic species are in opposite phase with respect to the phase of the laser field for the $1\times 10^{15}$ and $5\times 10^{15}\, Wcm^{-2}$ intensities.   
The amplitude of the variations of $<\!z\!>$ is higher for the \textit{ionic} species. This can be attributed to the stronger electron-electron repulsion and shielding effect of one electron on the other which is much higher in the \textit{ionic} species.

\begin{figure}[ht]
\begin{center}
\begin{tabular}{c}
\resizebox{90mm}{!}{\includegraphics{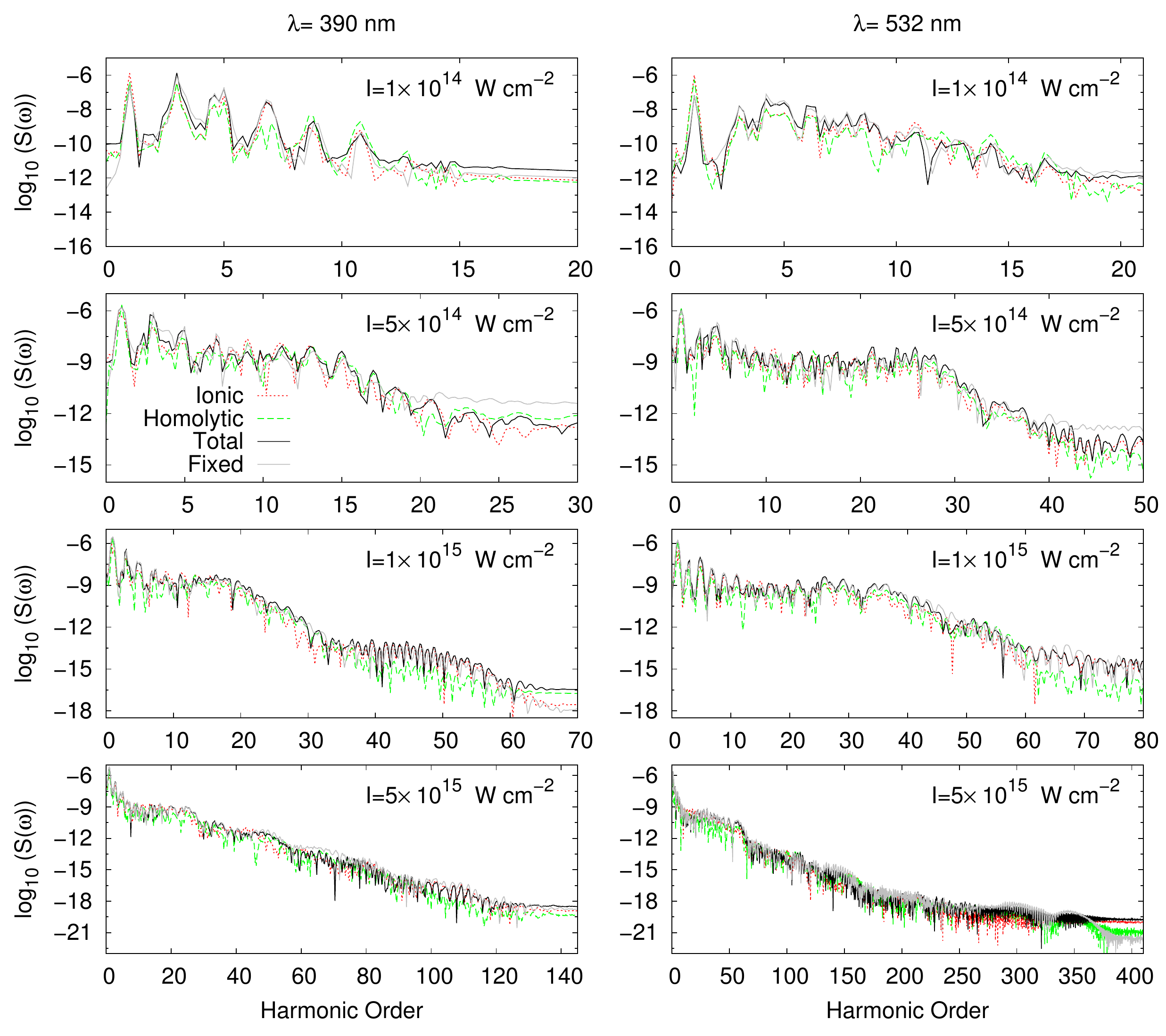}}
\end{tabular}
\caption{
\label{HHG} 
(color on line) The HHG spectra for freed nuclei, $S(\omega)$ as defined in Eq. (3), produced by the two-electron wavepacket evolution of the pre-ionization molecular H$_2$ (\textit{Total}) and transient \textit{ionic } $(H_\alpha^+-H_\beta^- $ $\&$ $ H_\alpha^--H_\beta^+)$ and \textit{homolytic}
 $(e_1 H_\alpha^+-H_\beta^+ e_2 $ $\&$ $ e_2 H_\alpha^+-H_\beta^+ e_1 )$ species over the whole period of interaction of the H$_2$ system with 8-cycle ultrafast intense laser pulses of $\lambda$= 390 nm (left) and 532 nm (right) wavelengths and at I=$1\times 10^{14}, 5\times10^{14}$, $1\times 10^{15}$ and $5\times 10^{15}\, Wcm^{-2}$ intensities. 
For comparison of the results obtained with the freed and fixed nuclei and to see the effect of the motion of nuclei, the HHG spectra obtained with the fixed nuclei are also presented. 
The same line style is used for all panels.		}
\end{center}
\end{figure}

The overall HHG power spectra obtained at the four laser intensities and the two wavelengths up to the end of the interaction period are plotted in Fig~\ref{HHG}. 
As can be seen from this figure, the HHG spectra follows a non-perturbative plateau behavior immediately after a steep descent at small harmonic orders which is due to the perturbative regime of interaction with the laser pulse, and before a cut-off. 
We recall that for sufficiently long laser pulses, only odd harmonics of the laser carrier angular frequency $\omega$ are observed in an isotropic gaseous medium having inversion symmetry \cite{Joachain_2012}.
In the case of atoms interacting with laser pulses comprising many optical cycles and having intensities such that non-dipole and relativistic effects can be neglected, the harmonic angular frequencies $\omega$ are only emitted at odd multiples of the laser angular frequency ($\omega = q\omega_0$, and $q =1,3,5, ...$) because of the inversion symmetry of the atom in the field.
When the driving laser pulse comprises of only a few optical cycles, the photons are emitted with a continuous distribution of frequencies, not at discrete harmonic frequencies only \cite{Joachain_2012}.
The laser fields in this work contain 8 cycles with the $sin^2$ shape carrier envelope. It can be seen in Fig.~\ref{HHG} that odd harmonics dominate in the low harmonic orders of the HHG spectra. In the higher harmonic orders in the HHG spectra, this structure gradually disappears and a specific order cannot be identified. 
Figure~\ref{HHG} also shows that this disordering of the HHG spectrum is intensified with the increase in the laser intensity. 
As can be seen in Figs.~\ref{norm} and \ref{<z>}, with increasing the laser intensity, population of H$_2$ is ionized completely earlier even before reaching the peak of the laser field envelop. 
So, the pre-ionized H$_2$ actually senses less optical cycles of the laser field and therefore the photons are emitted with a continuous distribution of frequencies.
Figure~\ref{HHG} also shows the HHG spectrum for the pre-ionization ionic and homolytic species.
In this figure, it is possible to compare the weights of the ionic and homolytic HHG spectra with each other and also with the overall H$_2$ spectrum.
 For example, it can be seen in Fig.~\ref{HHG} that the HHG spectrum of the ionic species dominates that of the homolytic species for the seventh harmonic order of the HHG spectrum produced by the
laser pulse of $1\times 10^{14}\, Wcm^{-2}$ intensity and 390 nm wavelength, and for 40-60 harmonic orders of the HHG spectrum produced by the $1\times 10^{15}\, Wcm^{-2}$ intensity and 390 nm laser pulses. 
In general, the behavior of the ionic and homolytic HHG spectra are very close. It can be justified as the HHG is due to the return of the electrons wavepacket towards the nuclei that mainly occurs near the common boundary of the homolytic and ionic species. 

\label{Effect of the motion of nuclei}
The HHG spectra obtained for the cases of the freed and fixed nuclei are also compared in Figure \ref{HHG}. 
The structures of the HHG of these two cases are different in details and are similar overally. 
Similar HHG structures shows that effect of the motion of nuclei is small.
This similarity would be predictable because the internuclear distance changes slightly in both wavelengths and all intensities (Fig.~\ref{internuclear_distance}).

\begin{figure}[h!]
\begin{center}
\begin{tabular}{c}
\resizebox{90mm}{!}{\includegraphics{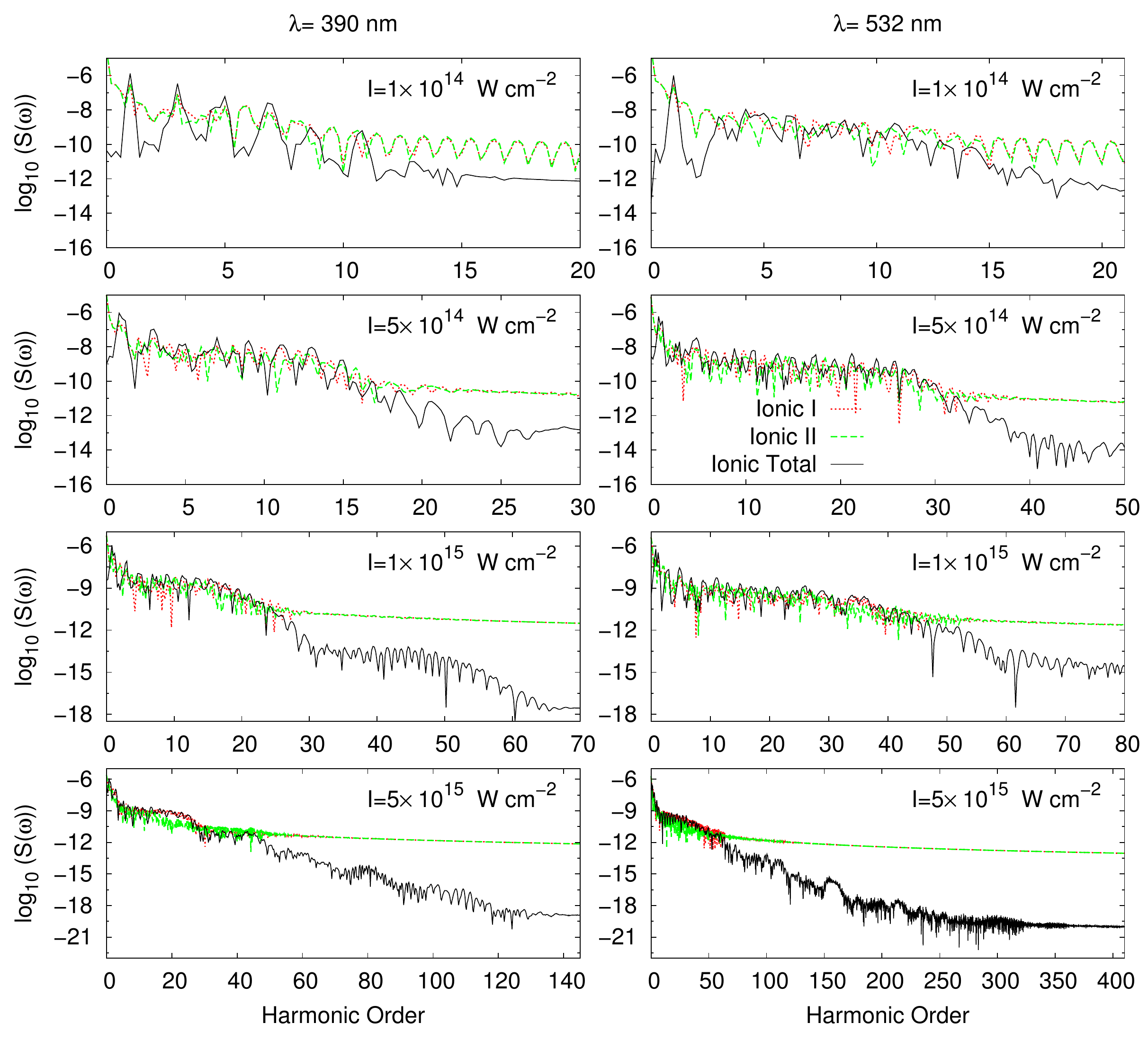}}
\end{tabular}
\caption{
\label{HHG_ionic} 
  (color on line) The same as Fig.~\ref{HHG}, but for different component ionic species; ionic I $(H_\alpha^--H_\beta^+)$, ionic II $(H_\alpha^+-H_\beta^-)$, and ionic Total $(H_\alpha^+-H_\beta^- $ $\&$ $ H_\alpha^--H_\beta^+)$.
}
\end{center}
\end{figure}

As we show in our previous article \cite{VSSBS_2012}, there are two different distinguishable regions  for the ionic part of the pre-ionized H$_2$. 
The HHG spectra produced by these parts are shown in Fig.~\ref{HHG_ionic}. 
It can be seen in this figure that to some extent at low laser intensity the even harmonic orders become weaker and odd harmonic orders become stronger. In all panels of Fig.~\ref{HHG_ionic}, it can be seen that at the high harmonic orders, the ionic I and II species partially cancel each other HHG, and the overall ionic HHG spectra falls down substantially.  
Also, it is interesting to find some locations in Fig.~\ref{HHG_ionic} at which the HHG of one of ionic region becomes dominant over that of the other ionic region. For example, at $5\times 10^{15}\, Wcm^{-2}$ laser intensity, this occurs at about 15-25 and 20-40 harmonic orders for 390 and 532 wavelengths, respectively.

\subsection{HHG radiation by the pre-ionization species of the hydrogen molecule}
\begin{figure*}[ht]
\begin{center}
\begin{tabular}{c}
\resizebox{180mm}{!}{\includegraphics{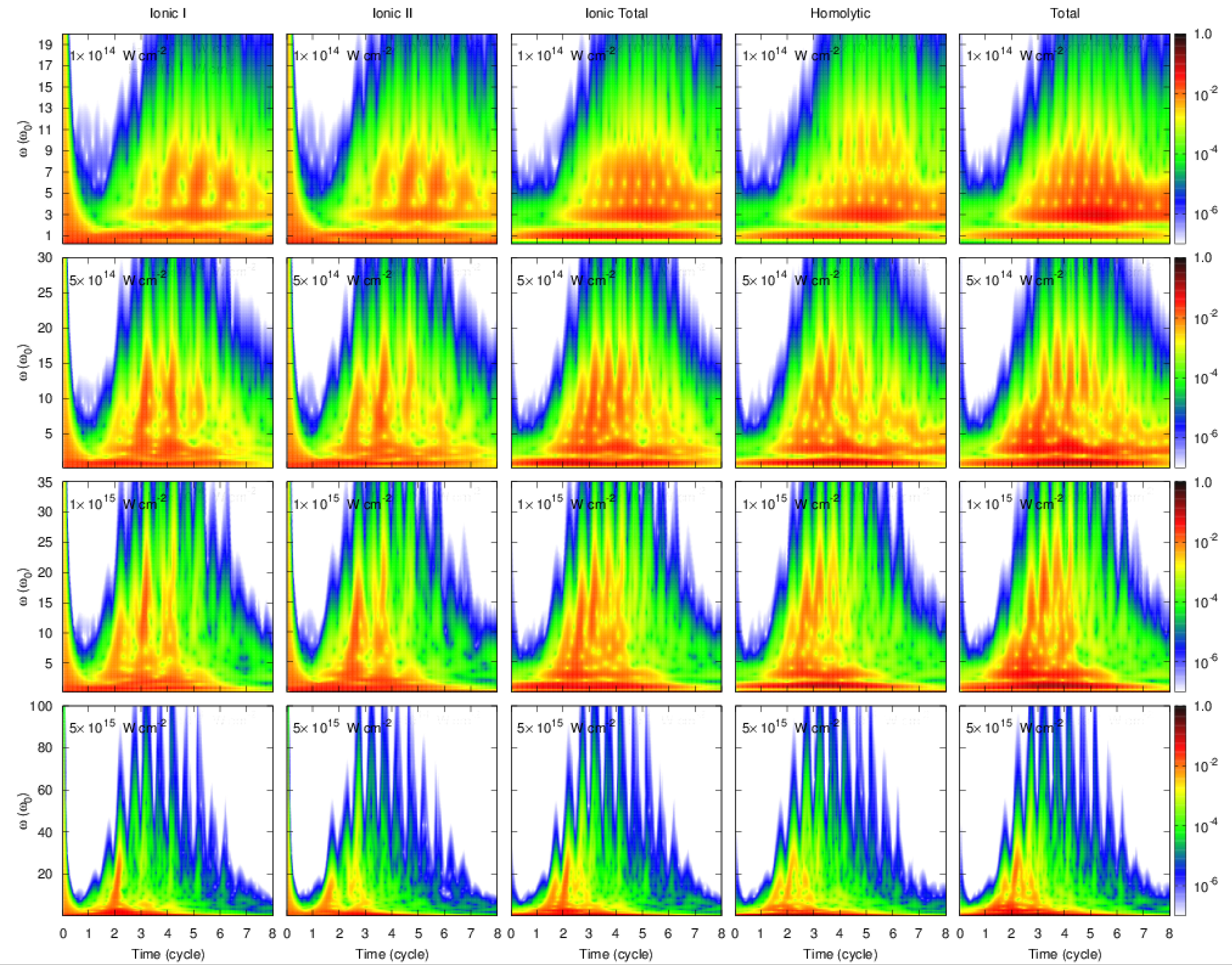}}
\end{tabular}
\caption{
\label{TP_390} 
  (color on line) The time-frequency profiles of the high harmonics produced by the two-electron wavepacket evolution obtained 
for the pre-ionization molecular H$_2$ (\textit{Total}),
\textit{homolytic} $(e_1 H_\alpha^+-H_\beta^+ e_2 $ $\&$ $ e_2 H_\alpha^+-H_\beta^+ e_1 )$ species and different components of the ionic species; 
ionic I $( H_\alpha^--H_\beta^+)$, 
ionic II $(H_\alpha^+-H_\beta^- $), and 
ionic Total $(H_\alpha^+-H_\beta^- $ $\&$ $ H_\alpha^--H_\beta^+)$ during the interaction of the H$_2$ system with 8-cycle ultrafast intense laser pulses of $\lambda$= 390 nm wavelength and at I=$1\times 10^{14}, 5\times10^{14}$, $1\times 10^{15}$ and $5\times 10^{15}\, Wcm^{-2}$ intensities.  
}
\end{center}
\end{figure*}

\begin{figure*}[ht]
\begin{center}
\begin{tabular}{c}
\resizebox{180mm}{!}{\includegraphics{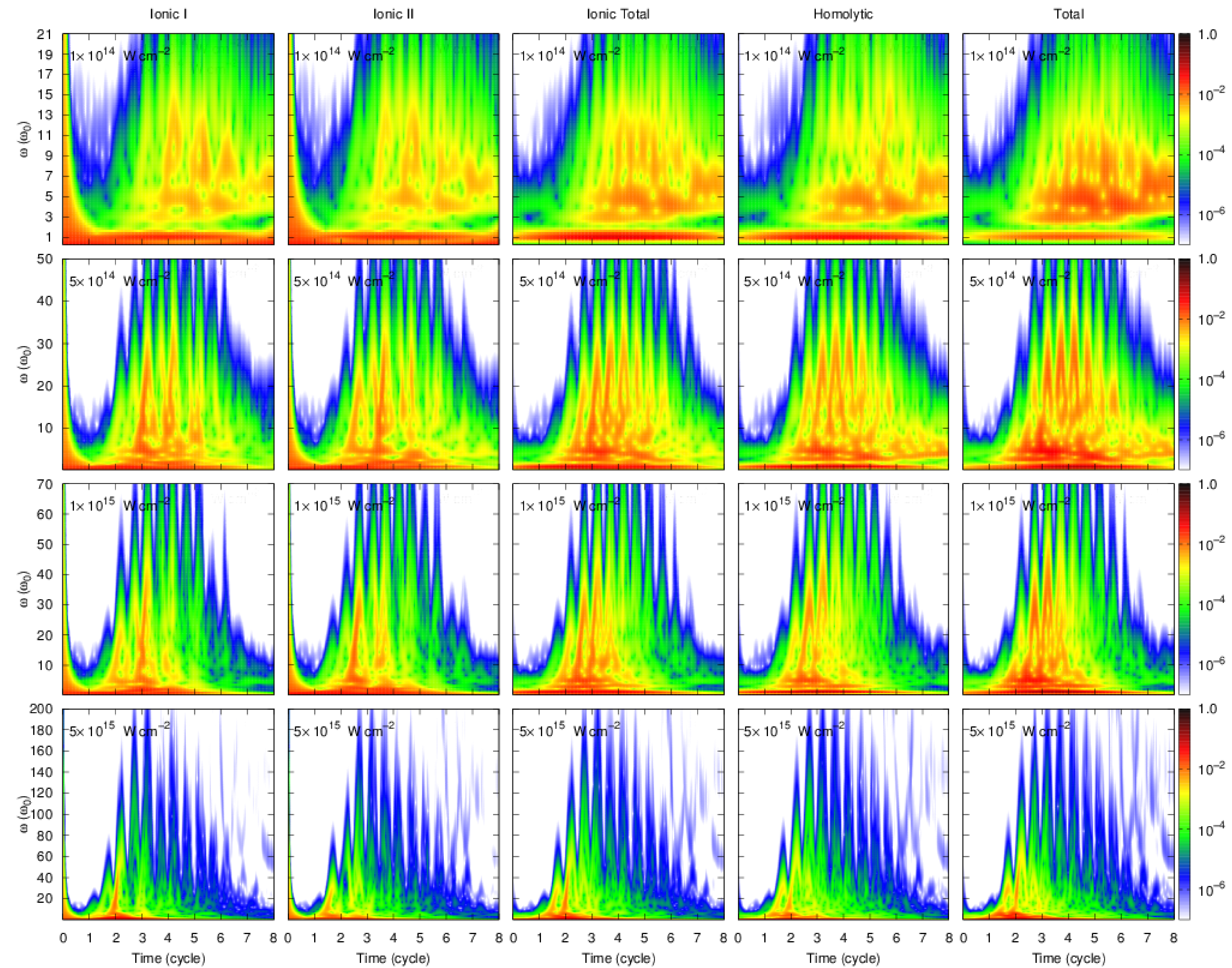}}
\end{tabular}
\caption{
\label{TP_532} 
(color on line) The same as Fig. 6, but for $\lambda$= 532 nm wavelength. 
}
\end{center}
\end{figure*}

Several processes result in the production of HHG by the pre-ionized hydrogen molecule in an intense laser field; simultaneous returning of the two electrons to the same nucleus, simultaneous returning of the two electrons to different nuclei, and returning of only one of the two electrons to one of the nuclei.
In the first process, two electrons simultaneously return to one of the nuclei, A or B. 
The locus of this process which occurs in the ionic I and II regions, is marked in Fig.~\ref{PIR} by points AA and BB. 
In this process, the motion of electrons is affected by the electron-electron repulsive interaction and
thus probability of this process and its weight in the HHG spectrum is very low. 
In the ionic I and II regions, both electrons return to the nuclei A and B, respectively.
In the second process, one electron returns to the nucleus A and the other returns to the nucleus B. 
This process occurs in the locus of points marked by AB in the homolytic III and IV regions in Fig.~\ref{PIR}. 
Generally, in the homolytic species, processes corresponding to the two electrons are indistinguishable. 
Since in the second process the two electrons return to different nuclei, they do not apply electrostatic Coulomb repulsion on each other considerably. Therefore, this process is more likely to occur than the first process.
The locus of the third process (returning of only one electron to one nucleus) is shows in Fig.~\ref{PIR} with lines 1, 2 , 3 and 4. 
Electron 2 approaches the nuclear system and reaches the nuclei A and B at lines 1 and 2, respectively. Similarly, in its approach towards the nuclear system, electron 1 reaches nuclei A and B at lines 3 and 4, respectively.
In the ionic I region, electron 2  is moved from the negative to the positive values of the $z_2$ coordinate following the inversion of the electric field of the laser pulse, and encounters with A nucleus at line 1 and is scattered. 
On its trajectory at line 2 in the homolytic region III, electron 2 is scattered by nucleus B.
Note that the simulation box is symmetric with respect to the diagonal D. Therefore, we only discuss the processes above this line \cite{VSSBS_2012}.
In our calculations, since the initial internuclear distance $R_0=2.13$ is small and does not change considerably during interaction with the field of the laser pulse (when there is considerable population in the pre-ionization H$_2$ region),
the internuclear distance is small in comparison with the dimension of the pre-ionization region (Fig.~\ref{internuclear_distance}) and the nuclei remain close to each other.
Therefore, each electron in its returning route is scattered mainly by one nucleus and the effect of the second nucleus on its scattering is small. 
For example, electron 2 in its displacement from negative to positive values of the $z_2$ coordinate is scattered strongly by the A nucleus (at line 1) and weakly by the B nucleus (at line 2).
This results in a weak HHG signal for the homolytic III region as opposed to the strong HHG signal produced by the ionic I region.
Simultaneously, above the diagonal line D, electron 1 moves from negative to positive values of the $z_1$ coordinate along lines 1 and 2. In this trajectory, electron 1 is scattered by the A nucleus (at line 3) strongly and by the B nucleus (at line 4) weakly resulting in a strong HHG signal in the homolytic III region and a weak HHG signal in the ionic II region.
In summary, in a typical half cycle of the laser field, we have a strong radiation in the ionic I region related to the crossing at line 1, a strong and a weak radiation in the homolytic III region related respectively to the crossing at lines 3 and 2, and finally a weak radiation in the ionic II region corresponding to the crossing of the electron trajectory at line 4.
In the next half cycle, a strong HHG radiation is produced by the ionic II region at line 4, a strong  and a weak HHG radiation is produced by the homolytic III region respectively at lines 2 and 3, and a weak HHG is produced by ionic I region at line 1.
Therefore, for each half cycle of the laser field, we see a strong and a weak radiation by the homolytic species and also a strong and a weak radiation by the ionic species.
Since, electrons are indistinguishable, a symmetric event is expected with respect to the permutation of the electrons 1 and 2  below the line D in Fig.~\ref{PIR}, so that the overall results for electrons 1 and 2 are totally indistinguishable.
If we study the ionic I and II regions individually, the ionic I region has a strong HHG radiation for a half cycle and a weak HHG radiation for the next half cycle, as shown in Figs.~\ref{TP_390} and \ref{TP_532}.
The ionic II region has inversely a weak and a strong HHG radiation, respectively for the same pair of consecutive half cycles.
We can see in Fig.~\ref{HHG} collectively that the HHG spectra of the homolytic and ionic species for different wavelengths and intensities are very similar and in some limited regions, one of them dominate the other.


\subsection{Time-frequency profiles of HHG}
The 2D color scale graphs of the time-frequency profiles of the high harmonics (TFP-HH) generated from the evolution of the two-electron H$_2$ system with freed nuclei are obtained using Morlet wavelet transform \cite{Ban_Che_Lu_2013,Uzer_2003} based on acceleration moment (Eq.~\ref{eq:4}), and presented in Figs.~\ref{TP_390} and \ref{TP_532}.
It can be seen from these figures that at all conditions of the laser pulse, the HHGs show bursting oscillative time profiles. 
At each burst, the electrons emit a continuous ${\omega}_n=n {\omega}_{\circ}$ frequency profile with decreasing intensity gradually with ${n}$ until reaching a cut-off. The overall heights of all bursts for the \textit{ionic} species matches the width of the plateau regime of the corresponding HHG spectra at each laser intensity and wavelength plotted in Fig.~\ref{HHG}.
In Fig.~\ref{TP_390}, for the $1\times 10^{14}\, Wcm^{-2}$ intensity, the bands of the odd harmonic orders of the laser pulse frequency are evident for both ionic and homolytic species. This structure gradually disappears for higher intensities. These odd-order bands appear weakly for the 532 nm wavelength at $1\times 10^{14}\, Wcm^{-2}$ intensity (Fig.~\ref{TP_532}) and disappears at higher intensities.
The corresponding time-frequency profiles of the homolytic and ionic species are very similar as shown in Figs.~\ref{TP_390} and \ref{TP_532}. 
As mentioned before, the HHG is mainly originated from electron scattering by nuclei, and the nuclei are located near the common boundaries of the ionic and homolytic species.
The radiation of the ionic and homolytic species are similar because of nearly symmetric distribution of the electron wavepacket over the pre-ionized H$_2$ and the change in the internuclear distance is not so large, as compares to the pre-ionized H$_2$ dimensions (when there is considerable population in this region), during irradiation of the pre-ionized hydrogen molecule by the intense laser field.
 
The presence of the well-known emissions corresponding to the short and long trajectories, and the highest harmonic orders (i.e. the cut off harmonics and beyond) \cite{Chirila__Lein_2010}, can clearly be seen in Figs.~\ref{TP_390} and \ref{TP_532}. These trajectories are more obvious for higher intensities of the laser field (in the tunnelling regime).

The Keldysh parameter \cite{Keldysh} of the intense laser pulses used in this study are reported in Table~\ref{Table}. It can be seen from this Table that the Keldysh parameter changes from 0.378 (in the tunnelling regime) for the longer wavelength (532 nm) and highest intensity ($5\times 10^{15}  Wcm^{-2}$) to 3.65 (in multi-photon excitation regime) for the shorter wavelength (390 nm) and lowest intensity ($1\times 10^{14} Wcm^{-2}$). When intensity is increased, the electronic wavepacket can distance farther from the nuclei in each cycle of the wavepacket evolution, and thus, outgoing and returning of the electrons occur more evidently.
It can be seen from Figs.~\ref{TP_390} and \ref{TP_532} that these trajectories disappear after a few cycles for higher intensities, because depletion of the population of the pre-ionized H$_2$ occurs earlier for higher intensities.

\begin{table}
\caption{\label{Table}The Keldysh parameter of the laser pulses used in this work.}
\centering
\begin{tabular}{|c|c|c|}
\cline{1-3}
 \multicolumn{1}{|c|}{}  & \multicolumn{2}{|c|}{Keldysh parameter}  \\\cline{2-3}
Intensity ($Wcm^{-2}$)    & 390 (nm) & 532 (nm)  \\  
 \hline
$1\times 10^{14}$      & 3.65    & 2.68      \\
$5\times 10^{14}$      & 1.63    & 1.20      \\
$1\times 10^{15}$      & 1.15    & 0.85      \\
$5\times 10^{15}$      & 0.52    & 0.38      \\
\hline
\end{tabular}
\end{table}

The TFP-HH of different regions of the ionic species are also demonstrated in Figs.~\ref{TP_390} and \ref{TP_532}.
These Figures show the sub-structures of TFP-HH of the ionic species.
It can be seen from these Figures that each of the ionic regions I and II has one burst for every consecutive pair of half cycles of the laser field, and together construct two bursts in each cycle for the total TFP-HH produced from the ionic species. This is in agreement with what is discussed in Sec. III-A.

The unrealistic initial strong signals appearing in Figs.~\ref{TP_390} and \ref{TP_532} for the regions of the ionic species are due to the limitation of the time duration over which the integration of the Morlet wavelet transformation is carried out. This limitation results in an artificial signal in the initial time limit of the TFP-HH. As can be seen in Figs.~\ref{TP_390} and \ref{TP_532}, these artefact signals are more significant for the ionic I and II regions, but are considerably diminished for the overall ionic species because of partial cancellation in the collective integration. These artefact signals for the ionic I and II regions and their cancellation for the ionic species are also observable at the high harmonics of the HHG spectra of Fig.~\ref{HHG_ionic}.

\section{Conclusion}
Hydrogen molecule, the simplest two electronic molecule, in a few-cycle intense laser field shows many interesting behavior. 
Upon the interaction with the intense laser field, the H$_2$ molecule is converted to a combination of the pre-ionized inert, and singly and doubly ionized species \cite{VSSBS_2012}.
 It is shown that the HHGs produced by the two ionic and homolytic pre-ionized species are similar at all laser pulse wavelengths and intensities used in this work. 
Effects of the laser pulse wavelength and intensity on the HHG production from the \textit{ionic} and \textit{homolytic} pre-ionization transient states are studied in detail. 
It is found that for the populated homolytic species in each half cycle of the laser pulse having enough strength, there is a pair of strong and weak HHG radiations produced simultaneously. 
Similarly, in each cycle, one of the ionic regions (I or II) has a strong radiation and the other ionic region (II or I) has a weak HHG radiation. In the next half cycle, the ionic region that had a strong HHG in the previous cycle shows a weak radiation, and vice versa for the other ionic region. Overally, similar to the homolytic species, the ionic species radiates a strong and a weak HHG spectra for each cycle of the laser field. 
The HHG spectra of the homolytic and ionic species are almost similar, except that in some limited parts of the HHG spectrum, one of them dominate the other.
The effect of the motion of nuclei is small and the HHG structures obtained with fixed and freed nuclei are overally very similar but different in details.
In the present study, we have focused on pre-ionization regions. In the next steps of this series of works, we will work on the ionized regions of the H$_2$ system, and investigate their contributions to the overall HHG structure \cite{SVANB_2013}.

\begin{acknowledgements}
We acknowledge Tarbiat Modares University, University of Isfahan and Laser-Plasma Research Institute of Shahid Beheshti University for financial supports and research facilities. We also thank Dr. Niknam for help in computing. 
\end{acknowledgements}

\section{References}
\bibliography{p7}
{}

\clearpage

\end{document}